\documentclass[a4paper,11pt]{article}
\usepackage{pos}
\usepackage{subfigure}
\usepackage{slashed}

\title{Exploring Background Contributions in $H \to Z \gamma$ Decay}

\author*[a]{Aliaksei Kachanovich}

\affiliation[a]{Service de Physique Théorique, CP225,\\
Université Libre de Bruxelles \\
Boulevard du Triomphe (Campus de la Plaine) \\
1050 Bruxelles
Belgium}

\emailAdd{aliaksei.kachanovich@ulb.be}

\abstract{The rare decay $H \to Z\gamma$ has been investigated by both the ATLAS and CMS Collaborations, with each reporting an excess in Run~2 in 2023 characterized by $\mu = 2.2 \pm 0.7$. This anomaly was initially attributed to possible modifications of the $HZ\gamma$ vertex. However, because the $H \to Z\gamma$ signal is reconstructed via the $H \to \ell\ell\gamma$ channel, background effects -particularly those from processes that mimic the same final state - may have been underestimated. In this work, we re-examine these backgrounds in detail and propose that the observed excess could be explained by an additional BSM-induced contribution. We present both an effective field theory framework and a UV-complete model that provide the necessary rates and a consistent interpretation of the data. The proposed model reproduces the newest ATLAS result, $\mu = 1.3^{+0.6}_{-0.5}$, obtained using a narrower dilepton invariant-mass window.}

\FullConference{European Physical Society Conference \\
on High Energy Physics \\
July 7-11, 2025 \\
Marseille
}

\begin{document}
\maketitle

\section{Introduction}
The Standard Model (SM) is extremely successful but remains incomplete. Significant efforts are devoted to searching for deviations from the SM. Following the discovery of the Brout–Englert–Higgs boson \cite{ATLAS:2012yve,CMS:2012qbp} (commonly referred to as the Higgs boson), many of its properties are still not fully understood. Recent results from the ATLAS and CMS collaborations provide the first evidence for the rare decay $H \rightarrow Z\gamma$ \cite{ATLAS:2020qcv,CMS:2022ahq}. The combined measurements from both experiments yield a branching fraction of $\text{B}_{\rm obs} = (3.4 \pm 1.1) \times 10^{-3}$, which exceeds the SM prediction by a factor of $(2.2 \pm 0.7)$ \cite{ATLAS:2023yqk}. The discrepancy corresponds to a statistical significance of only $1.9\sigma$. This result has motivated numerous studies aiming to provide an explanation (e.g. \cite{Barducci:2023zml,Boto:2023bpg,Das:2024tfe,Cheung:2024kml,Hu:2024slu,Hernandez-Juarez:2024pty,Zhang:2024yqw,Arhrib:2024wjj,Israr:2024ubp,Mantzaropoulos:2024vpe,Sang:2024vqk}). Most of these works attempted to explain this excess as a modification of the $HZ\gamma$ vertex.

The process $H \to Z\gamma$ is reconstructed from the $\ell\ell\gamma$ final state ($\ell = e, \mu$). The same final state also receives contributions from $H \to \gamma\gamma$, box diagrams, and $H \to \mu\mu\gamma$ via the Yukawa coupling~\cite{Kachanovich:2020xyg,Kachanovich:2021pvx,Corbett:2021iob,Chen:2021ibm,Ahmed:2023vyl,Hue:2023tdz,VanOn:2021myp,Sun:2013cba,Phan:2021xwc,Phan:2021ovj,Kachanovich:2024vpt,Abbasabadi:1996ze,Chen:2012ju,Dicus:2013ycd,Passarino:2013nka,Han:2017yhy,Nisandzic:2025raw}.  
A dilepton invariant-mass cut of $m_{\ell\ell} > 50~\mathrm{GeV}$~\cite{CMS:2022ahq,ATLAS:2023yqk} is applied to suppress the $H \to \gamma\gamma$ background, which reduces but does not completely remove it (see Fig.~\ref{fig:Rescale}).

Together with J.~Kimus, S.~Lowette, and M.H.G.~Tytgat, we propose a new approach to addressing the excess by introducing an additional background from BSM processes in Higgs decays~\cite{Kachanovich:2025cxz}. By applying a narrower dilepton invariant-mass window centered on the $Z$ peak ($|m_{\ell\ell} - m_Z| < 10~\text{GeV}$), we obtain a signal strength of $\mu = 1.39$ (see Tab.~\ref{tab:sample}). This result is also consistent with the recent ATLAS measurement~\cite{ATLAS:2025aip}, which used a similar invariant-mass window around the $Z$ peak and reported a signal strength closer to the SM expectation, $\mu = 1.3^{+0.6}_{-0.5}$. It is important to mention that, an additional 6.4\% contribution to the measurement in the dilepton invariant-mass window between 50 and 125~GeV arises from other SM processes.

\section{The Standard Model}
The SM contribution to $H \to \ell \ell \gamma$ is represented by four different sub-processes: the tree-level decay with Bremsstrahlung, the one-loop contribution with the $\gamma \gamma$ sub-process, the direct box-diagram coupling, and the one-loop contribution with $Z\gamma$, which are of particular interest here. For electrons, the tree-level contribution can be neglected, whereas in the case of muons it accounts for up to $70\%$\footnote{Depending on the kinematic cuts.} of the total decay rate. All one-loop contributions can be expressed in a general form as

\begin{eqnarray}
    \mathcal{M}_{\text{SM,loop}} &=& \nonumber \left[q_{\mu} p_{1} \cdot \varepsilon^{*}(q) - \varepsilon^{*}_\mu(q) \, q\cdot p_{1}\right] \bar{u}(p_{2}) \big( a_{1} \gamma^{\mu} P_{R} + b_{1} \gamma^{\mu} P_{L} \big) v(p_{1}) \\
    &+& \left[q_{\mu} p_{2} \cdot \varepsilon^{*}(q) - \varepsilon^{*}_\mu(q) \, q\cdot p_{2}\right] \bar{u}(p_{2})\big( a_{2} \gamma^{\mu} P_{R} + b_{2} \gamma^{\mu} P_{L} \big) v(p_{1}) \,,
    \label{eq:SMloop}
\end{eqnarray}
here the four-momenta of the leptons and photon are denoted by $p_{1}$, $p_{2}$, and $q$, respectively. The coefficients $a_{1(2)}$ and $b_{1(2)}$ depend on the Mandelstam variables, defined as $u = (p_2 + q)^2$, $t = (p_1 + q)^2$, and $s = (p_1 + p_2)^2$. The coefficients $a_{1(2)}$ ($b_{1(2)}$) are symmetric under the interchange of the variables $u$ and $t$. For explicit expressions of $a_{1(2)}$ and $b_{1(2)}$, see \cite{Kachanovich:2020xyg}.

All contributions can be separated into a resonant part, which can be associated with the $H \to Z\gamma$ process, and a non-resonant part, which can be defined by splitting the coefficients in Eq.~\ref{eq:SMloop} into two subsets (see details in \cite{Kachanovich:2021pvx}): 

\begin{center}
\begin{figure}[t]
\centering
{\includegraphics[width=0.68\textwidth]{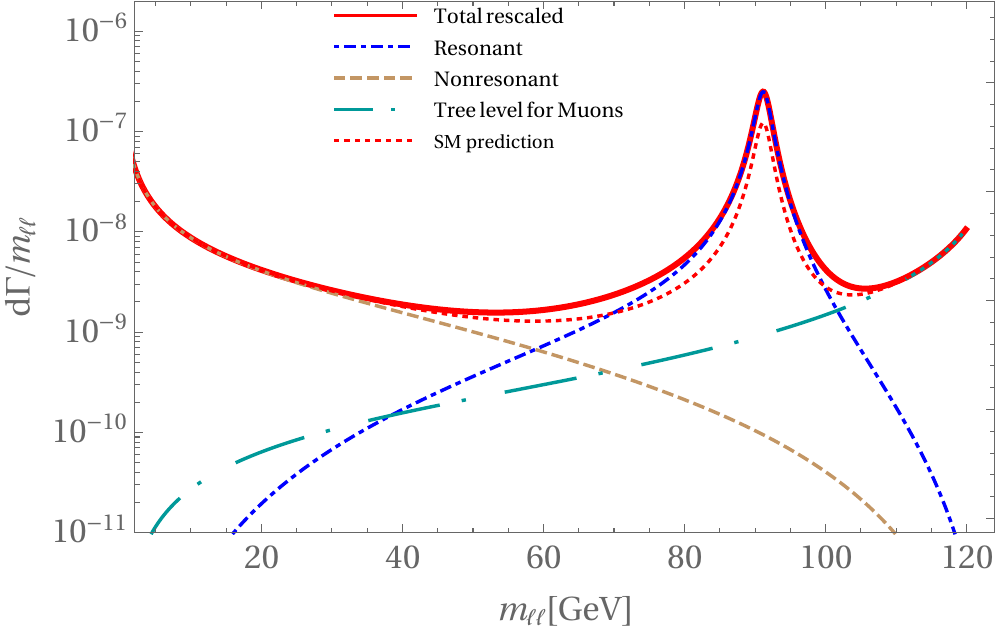}}
	\caption{
        Solid red: total with modified $HZ\gamma$; blue dash–dotted: rescaled resonant SM; short–dashed red: SM; brown dashed: non-resonant; long dash–dotted green: tree level.  No kinematic cuts were implemented.} 
	\label{fig:Rescale}
\end{figure}
\end{center}

\begin{eqnarray}
    a_{1(2)} (s,t) =  a_{1(2)}^{res}(s) + a_{1(2)}^{nr}(s,t) \,, 
\end{eqnarray}
with  
\begin{eqnarray}
     a_{1(2)}^{res}(s) &\equiv& \frac{\alpha (m_{Z}^2)}{s - m_{Z}^2 + i m_{Z} \Gamma_Z}, \qquad \mbox{\rm and} \qquad  
     a_{1(2)}^{nr}(s,t) \equiv \tilde{a}_{1(2)} (s, t) + \frac{\alpha(s) - \alpha (m_{Z}^2)}{s - m_{Z}^2 + i m_{Z} \Gamma_Z}\,,
\end{eqnarray}  
where $\alpha(s)$ denotes the symmetric part of the coefficients $a_{1(2)}$, since it is the same for both. A similar definition symmetric part $\beta(s)$ applies to the form factors $b_{1(2)}$. The original form factor can then be rewritten as  
\begin{eqnarray}
    a_{1(2)}(s,t) = \tilde{a}_{1(2)}(s,t) + \frac{\alpha (s)}{s - m_{Z}^2 + i m_{Z} \Gamma_Z}\,.
\end{eqnarray}  
with an analogous expression for $b_{1(2)}(s,t)$.

The resonant contribution depends solely on the dilepton squared invariant mass, denoted by the variable $s$. It is also important to emphasize that one cannot isolate a subset of diagrams to obtain the resonant part, since gauge-dependent contributions cancel only after summing over all diagrams (full set of diagrams one can find in \cite{Kachanovich:2020dah}).

By rescaling the $H \to Z \gamma$ vertex one can explain excess in $H \to \ell \ell \gamma$ in particular kinematic region (see Fig.~\ref{fig:Rescale}).

\section{New Physics}
\subsection{Effective Field Theory}\label{sec.EFT}

\begin{figure}[]
	\begin{center}
		\subfigure[t][]{\includegraphics[width=0.49\textwidth]{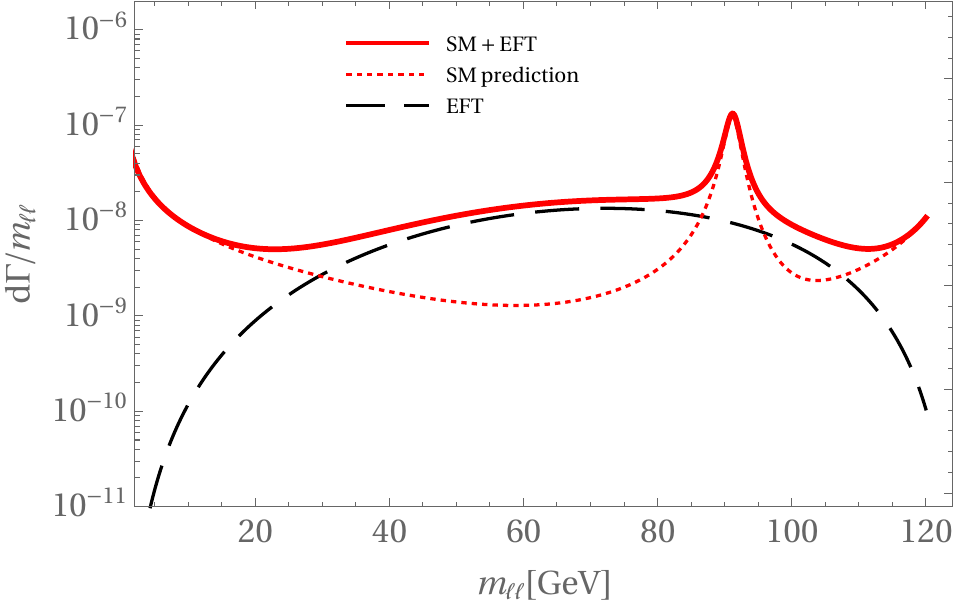}}
		\hspace{.1cm}
  	\subfigure[t][]{\includegraphics[width=0.49\textwidth]{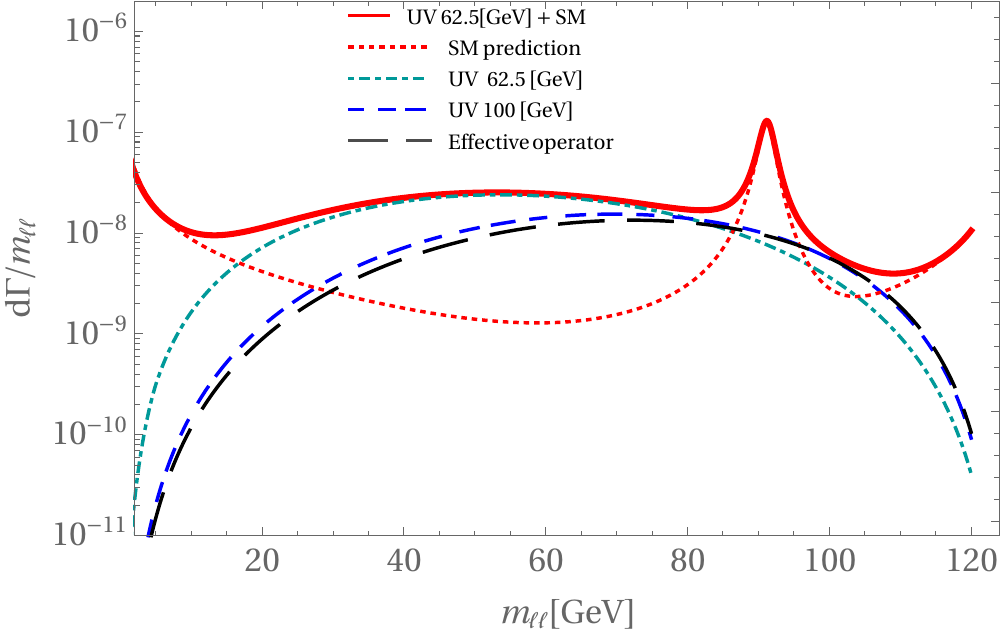}}
         \end{center}
	\caption{(a) Contribution of the EFT operator          
    (Eq.~\ref{eq:Lag_EFT}) to $H \to \ell\ell\gamma$ with $\Lambda_R=260$~GeV, chosen to match the observed excess over the SM. (b) UV-complete benchmarks: $M=m_S=m_{\Psi}=62.5$~GeV (dot–dashed green; solid red) and $M=100$~GeV. The black long-dashed curve shows the EFT result from (a). No kinematic cuts were implemented.}
	\label{fig:BSM_tot}
\end{figure}

The most common explanation for the excess is a modification of the $HZ\gamma$ vertex (e.g., \cite{Barducci:2023zml,Boto:2023bpg,Das:2024tfe,Hu:2024slu,Hernandez-Juarez:2024pty,Zhang:2024yqw,Arhrib:2024wjj,Israr:2024ubp,Mantzaropoulos:2024vpe,Sang:2024vqk}). As an alternative, we propose the presence of an additional background. A general way to incorporate BSM contributions is through an Effective Field Theory (EFT) framework. Although several operators can contribute to the process $H \to \ell \ell \gamma$, for simplicity we focus on a single dimension-8 operator:

\begin{equation}\label{eq:Lag_EFT}
   {\cal L}_{\rm eff} \supset {g'\over \Lambda_R^4} \vert \Phi  \vert^2 \partial_{\nu} (\bar \ell_{R} \gamma_{\mu} \ell_R) B^{\mu \nu} \,,
\end{equation}

\noindent where $\Phi$ is the SM Higgs doublet, $\ell_R$ denotes the $SU(2)$-singlet right-handed lepton spinors, $B^{\mu\nu}$ is the hypercharge field strength, and $g' = e/\cos\theta_W$ is the SM hypercharge coupling. The underlying dimension-6 structure, $\partial_{\nu} (\bar{\ell}_R \gamma_{\mu} \ell_R) B^{\mu \nu}$, vanishes for on-shell fields; hence, the factor $\lvert \Phi \rvert^2$ is essential.

\subsection{UV-complete model}

There are many possible ways to construct a UV-complete model that potentially reproduces the background corresponding to the EFT scenario presented in Sec.~\ref{sec.EFT}. Here, we consider one specific UV-complete model~\cite{Toma:2013bka,Giacchino:2013bta}, in which the scalar particle $S$\footnote{It is the singlet under the SM.} can serve as a dark matter candidate~\cite{Silveira:1985rk,McDonald:1993ex,Burgess:2000yq}:
 
\begin{equation}\label{eq:LagUV}
    \mathcal{L}_{UV} \supset \frac{1}{2} \partial_\mu S \partial^\mu S - \frac{1}{2} m_S^{2} S^{2} + \bar{\Psi} (i \slashed{D} - m_{\Psi})\Psi - \sum_{\ell} (y_{\ell} S \bar{\Psi} \ell_{R} + h.c.) - \frac{\lambda_{hs}}{2} S^2 |H|^{2} 
\end{equation}

\begin{figure}[ht]
	\begin{center}
		\subfigure[t][]{\includegraphics[width=0.23\textwidth]{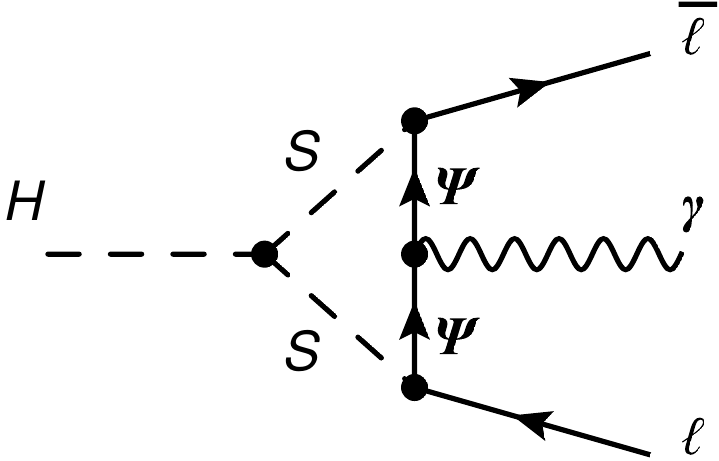}}
		\hspace{.6cm}
  	\subfigure[t][]{\includegraphics[width=0.23\textwidth]{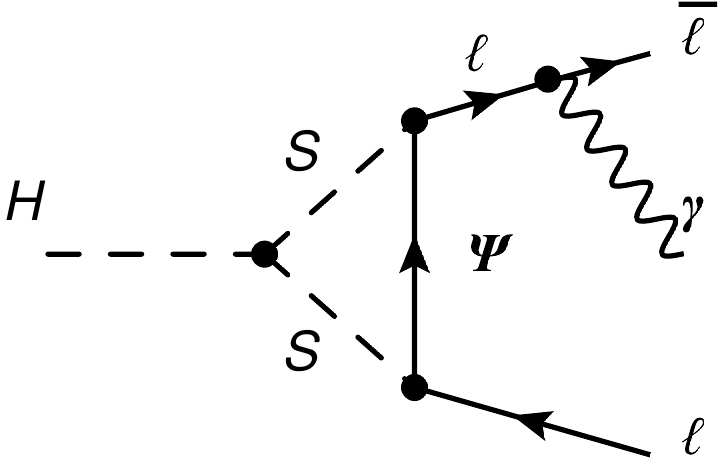}}
   		\hspace{.6cm}
   		\subfigure[t][]{\includegraphics[width=0.23\textwidth]{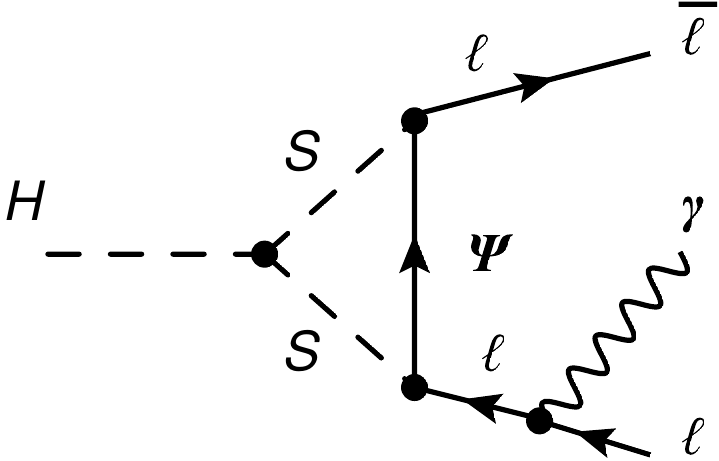}}
         \end{center}
	\caption{Amplitudes for the UV-complete model. Bremsstrahlung from final-state leptons (b, c) is loop- and chirality-suppressed, while the virtual fermion emission (a) reproduces the structure of the effective operator in Eq.~\ref{eq:Lag_EFT}.  }
	\label{fig:Loop_diagram}
\end{figure}

\noindent The scalar $S$ interacts with the SM through the Higgs portal and couples to right-handed leptons via a Yukawa interaction with a vector-like fermion $\Psi$ carrying hypercharge $Y = Q = -1$. Additional diagrams that provides this model are represented on Fig.\ref{fig:Loop_diagram}.

With this Lagrangian, the process $H \to \ell \ell \gamma$ receives an additional contribution from three extra diagrams (Fig.~\ref{fig:Loop_diagram}), which were evaluated using \emph{FeynCalc}~\cite{Shtabovenko:2016sxi, Shtabovenko:2020gxv, Shtabovenko:2023idz}.

\section{Phenomenology}
\subsection{General results}

To extract $H \to Z\gamma$ from $H \to \ell\ell\gamma$, one must account for additional backgrounds: $H \to \gamma\gamma$, non-resonant box-diagram $\ell\ell\gamma$ production, and off-shell contributions. Properly chosen kinematic cuts can reduce - but not completely remove - backgrounds from other SM processes. Comparing the ``pure'' $H \to Z\gamma$ result ($\mu = 2.20$) with the reconstructed one ($\mu = 2.06$) shows a $\sim 6.4\%$ decrease. If SM backgrounds are correctly modeled, any additional excess would have to originate from BSM contributions. 

In the new results~\cite{ATLAS:2025aip}, the ATLAS Collaboration adopted a narrower dilepton invariant-mass window, $|m_{\ell\ell} - m_Z|<10~\text{GeV}$, and obtained a signal strength closer to the SM expectation, $\mu = 1.3^{+0.6}_{-0.5}$, which corresponds precisely to the result presented in~\cite{Kachanovich:2025cxz} (see Fig.~\ref{fig:BSM_tot} and Tab.~\ref{tab:sample}, result \# 7, columns $Br_{\rm EFT}/Br_{\rm SM}$ and $Br_{\rm UV}/Br_{\rm SM}$).

The event yields reported in~\cite{ATLAS:2020qcv,ATLAS:2023yqk,ATLAS:2025aip,CMS:2022ahq} can be reproduced within the EFT description (Eq.~\ref{eq:Lag_EFT}) using the operator normalized at the scale $\Lambda = 260~\text{GeV}$. For the UV-complete theory, the parameter space contains four parameters, and many combinations can reproduce the required background rate. We focus on two representative mass points: $m_{\Psi}=m_{S}=62.5~\text{GeV}$ (i.e., $m_H/2$) and $m_{\Psi}=m_{S}=100~\text{GeV}$. In both scenarios we fix the Higgs–scalar coupling to $\lambda_{hs}=0.26$. The corresponding lepton Yukawa couplings needed to match the rate are $y_{\ell}=1.66$ for the $62.5~\text{GeV}$ point  for the $100~\text{GeV}$ point the product of couplings is already large $\lambda_{hs} y_{\ell}^2 = 28.1$.

\begin{table}[ht!]
\centering
\begin{tabular}{|c|c|c|c|c|c|c|c|c|}
\hline
\,\#\,& {Cuts} & $m_{\ell \ell}^{min}$[GeV] & $m_{\ell \ell}^{max}$[GeV]  & $\Gamma_{\rm tot}^{\rm SM}$ [keV] & $\Gamma^{\rm SM}_{\rm tree}$[keV] & {$\frac{Br_{\rm resc}}{Br_{\rm SM}}$} & {$\frac{Br_{\rm EFT}}{Br_{\rm SM}}$} & {$\frac{Br_{\rm UV}}{Br_{\rm SM}}$}\\
\hline
1 & \, None \, & 50 & 125  & 0.768 & 0.287 & 1.67 & 1.86 & 2.07\\
\hline
2 & None & 50 & 100  & 0.504 & 0.028 & 2.01 & 2.21 & 2.57\\
\hline
3 & CMS & 40 & 125  & 0.455 & 0.011 & 2.04 & 2.10 & 2.13 \\
\hline\hline
\textbf{4} & \textbf{CMS} & \textbf{50} & \textbf{125} & \textbf{0.451} & \textbf{0.011}  & \textbf{2.06} & \textbf{2.06} & \textbf{2.06} \\
\hline\hline
5 & CMS & 70 & 125   & 0.440 & 0.011  & 2.07 & 1.80 & 1.71\\
\hline
6 & CMS & 70 & 100   & 0.432 & 0.006  & 2.08 & 1.74 & 1.68 \\
\hline
7 & CMS & 80 & 100   & 0.416 & 0.005  & 2.09 & 1.48 & 1.39\\
\hline
\end{tabular}
\caption{Signal strengths for selected cut choices. The first two rows apply only the $m_{\ell\ell}$ cuts. All values are obtained by integrating over the Mandelstam $t$ variable. CMS-like cuts: $E_\gamma \ge 15$ GeV, $E_1 \ge 7$ GeV, $E_2 \ge 25$ GeV, and $t_{\min},u_{\min} \ge (0.1,m_H)^2$. Columns: $Br_{\rm resc}/Br_{\rm SM}$ - Z-peak strength with the effective $HZ\gamma$ coupling rescaled by $\sqrt{2.11}$ to match $\text{Br}_{\mathrm{obs}}=(3.4\pm1.1)\times10^{-3}$; $Br_{\rm EFT}/Br_{\rm SM}$ -EFT result (Eq.~\ref{eq:Lag_EFT}); $Br_{\rm UV}/Br_{\rm SM}$ - UV model (Eq.~\ref{eq:LagUV}) with $m_S = m_{\Psi}=62.5$ GeV}
\label{tab:sample}
\end{table}

\subsection{Phenomenological restriction}

The introduction of the background in Eqs.~\ref{eq:Lag_EFT} and \ref{eq:LagUV} does not contradict the main phenomenological constraints. The UV-complete model contributes to the anomalous muon magnetic moment and is consistent with the $(g\!-\!2)_{\ell}$ measurement for both mass benchmarks $m_{\Psi} = m_{S} = 62.5 (100)\,\text{GeV}$. According to the new results \cite{Aliberti:2025beg} the corresponding constraint for coupling constants is $\lambda_{hs}\, y_{\ell}^{2} < 0.48 (18.7)$.

The vector-like lepton $F \sim (1,1,-1)$ is relatively light and couples to both the photon and the $Z$ boson, potentially contributing to the oblique parameters~\cite{Peskin:1990zt}. Since our benchmark masses are comparable to $m_Z$ ($m_{\Psi} \lesssim m_Z$), the standard $S, T, U$ formalism is insufficient, and we instead use the extended set ($S, \ldots, X$) from Ref.~\cite{Maksymyk:1993zm}. Using the results of Ref.~\cite{Albergaria:2023nby} for general vector-like electroweak representations (see also~\cite{Moreno:1991mf}), we find
\begin{equation}
    S = -U = -0.013\,(-0.0042), \quad  
    V = -0.026\,(-0.0065), \quad  
    X = 0.014\,(0.0045)
\end{equation}
for our benchmarks $m_{\Psi} = 62.5\,(100)$~GeV. We find that the oblique corrections are well below the $1\sigma$ experimental limits:

\begin{align}
\nonumber \left.\frac{\Delta \Gamma}{\Gamma}\right|_{\mathrm{inv}} &= 0.0002\, (0.00005) < 0.003, 
& \left.\frac{\Delta \Gamma}{\Gamma}\right|_{\ell\ell} &= 0.0002\, (0.00006) < 0.001, \\
\Delta m_W &= 0.0082\, (0.0026)\ \mathrm{GeV} < 0.013\ \mathrm{GeV}.
\end{align}

\noindent For details, see Ref.~\cite{Kachanovich:2025cxz}.

The particles $F$ and $S$ are assumed to be nearly mass-degenerate, with $m_\Psi \gtrsim m_S$. The mass difference is sufficient for decays into electrons and muons but not into any heavier particles. Since the Yukawa coupling is not negligible, the decay $\overset{\scriptscriptstyle(-)}{F} \to S + \overset{\scriptscriptstyle(-)}{\ell}$ produces soft leptons that are invisible to LHC detectors. Consequently, $F\bar{F}$ production appears as missing energy, $pp \to \text{jets} + \slashed{E}$. From missing energy searches, the mass bound for the charged fermion particle is
$m_{\Psi} \gtrsim 67~\mathrm{GeV}$ (within $5\%$), which is borderline for
$m_{\Psi} \gtrsim m_S = 62.5~\mathrm{GeV}$, while
$m_{\Psi} \gtrsim m_S = 100~\mathrm{GeV}$ is safe.


In the context of dark matter phenomenology, we note that although the scalar $S$ could in principle serve as a DM candidate, its coupling to the SM through the Higgs portal leads to strong constraints: for the benchmark $m_S = 100$ GeV, thermal freeze-out would require $\lambda_{hs} \gtrsim 2.2$, whereas a dominant DM component would correspond to only $\lambda_{hs} \simeq 0.04$, implying that a stable $S$ would constitute only a subdominant fraction $f_S \approx (0.04/2.2)^2 \sim 3\times10^{-4}$. This mass–coupling requirement is generally excluded by direct-detection constraints, unless the lighter benchmark near the Higgs resonance ($m_S \approx m_H/2$) is considered, where suppression effects can allow compatibility. However, these arguments typically ignore the additional Yukawa interactions mediated by $F$, enabling $t$-channel DM processes that substantially enrich the dark-matter phenomenology -  a detailed exploration lies outside the scope of this work.

\section{Conclusions}

In this work, we presented a new idea to explain the excess $\mu = 2.2 \pm 0.7$ reported by both the ATLAS and CMS collaborations in the measurement of $H \to Z \gamma$ by introducing an additional background to $H \to \ell \ell \gamma$. Our approach reproduces the new ATLAS result~\cite{ATLAS:2025aip}, $\mu = 1.3^{+0.6}_{-0.5}$. BSM effects are considered both through an effective operator (Eq.~\ref{eq:Lag_EFT}) and a simplified UV model (Eq.~\ref{eq:LagUV}) with a scale close to the electroweak scale, also motivated by the dark matter problem.


 The differential decay rate for the rescaled $HZ\gamma$ vertex (Fig.~\ref{fig:Rescale}) is compared with the background contributions from the EFT and UV-complete models (Fig.~\ref{fig:BSM_tot}). The resulting signal shapes differ significantly between the models, highlighting the need for further experimental studies across dilepton mass bins $m_{\ell \ell}$. Different scenarios with kinematic cuts on $m_{\ell \ell}$, including experimental selections, are summarized in Tab.~\ref{tab:sample}.

It is deemed unlikely that new physics significantly contributes to $H \rightarrow \ell^+\ell^- \gamma$. The UV model considered is ad hoc and fine-tuned, relying on several unnatural assumptions, including a low new physics scale, a compressed mass spectrum ($m_S \approx m_F$), and identical Yukawa couplings for electrons and muons. Nevertheless, other experimental constraints are considered, along with the possibility that one of the new particles could contribute to dark matter. The model should remain relevant for future analyses, even if the excess disappears with the same cuts.

In this work, we investigated various constraints from $(g-2)_{\ell}$ and collider measurements, electroweak precision tests, and dark matter observations, all of which are consistent with our models. 

\section*{Acknowledgements}
I thank my colleagues Jean Kimus, Steven Lowette, and Michel H.G. Tytgat for the extremely enjoyable collaboration on this work. I am also grateful to Giorgio Arcadi, Debtosh Chowdhury, and Laura Lopez Honorez for useful discussions, as well as to Ivan Nišandžić, whose previous collaboration provided the idea for this work. In addition, I thank Vladyslav Shtabovenko for all kinds of support with \emph{FeynCalc} and \emph{FeynHelpers}. This research was supported by FRS/FNRS, FRIA, the BLU-ULB Brussels Laboratory of the Universe, and the IISN convention No. 4.4503.15.

\bibliographystyle{apsrev4-1}

\bibliography{main}

\begin{thebibliography}{49}%
\makeatletter
\providecommand \@ifxundefined [1]{%
 \@ifx{#1\undefined}
}%
\providecommand \@ifnum [1]{%
 \ifnum #1\expandafter \@firstoftwo
 \else \expandafter \@secondoftwo
 \fi
}%
\providecommand \@ifx [1]{%
 \ifx #1\expandafter \@firstoftwo
 \else \expandafter \@secondoftwo
 \fi
}%
\providecommand \natexlab [1]{#1}%
\providecommand \enquote  [1]{``#1''}%
\providecommand \bibnamefont  [1]{#1}%
\providecommand \bibfnamefont [1]{#1}%
\providecommand \citenamefont [1]{#1}%
\providecommand \href@noop [0]{\@secondoftwo}%
\providecommand \href [0]{\begingroup \@sanitize@url \@href}%
\providecommand \@href[1]{\@@startlink{#1}\@@href}%
\providecommand \@@href[1]{\endgroup#1\@@endlink}%
\providecommand \@sanitize@url [0]{\catcode `\\12\catcode `\$12\catcode `\&12\catcode `\#12\catcode `\^12\catcode `\_12\catcode `\%12\relax}%
\providecommand \@@startlink[1]{}%
\providecommand \@@endlink[0]{}%
\providecommand \url  [0]{\begingroup\@sanitize@url \@url }%
\providecommand \@url [1]{\endgroup\@href {#1}{\urlprefix }}%
\providecommand \urlprefix  [0]{URL }%
\providecommand \Eprint [0]{\href }%
\providecommand \doibase [0]{http://dx.doi.org/}%
\providecommand \selectlanguage [0]{\@gobble}%
\providecommand \bibinfo  [0]{\@secondoftwo}%
\providecommand \bibfield  [0]{\@secondoftwo}%
\providecommand \translation [1]{[#1]}%
\providecommand \BibitemOpen [0]{}%
\providecommand \bibitemStop [0]{}%
\providecommand \bibitemNoStop [0]{.\EOS\space}%
\providecommand \EOS [0]{\spacefactor3000\relax}%
\providecommand \BibitemShut  [1]{\csname bibitem#1\endcsname}%
\let\auto@bib@innerbib\@empty
\bibitem [{\citenamefont {Aad}\ \emph {et~al.}(2012)\citenamefont {Aad} \emph {et~al.}}]{ATLAS:2012yve}%
  \BibitemOpen
  \bibfield  {author} {\bibinfo {author} {\bibfnamefont {G.}~\bibnamefont {Aad}} \emph {et~al.} (\bibinfo {collaboration} {ATLAS}),\ }\href {\doibase 10.1016/j.physletb.2012.08.020} {\bibfield  {journal} {\bibinfo  {journal} {Phys. Lett. B}\ }\textbf {\bibinfo {volume} {716}},\ \bibinfo {pages} {1} (\bibinfo {year} {2012})},\ \Eprint {http://arxiv.org/abs/1207.7214} {arXiv:1207.7214 [hep-ex]} \BibitemShut {NoStop}%
\bibitem [{\citenamefont {Chatrchyan}\ \emph {et~al.}(2012)\citenamefont {Chatrchyan} \emph {et~al.}}]{CMS:2012qbp}%
  \BibitemOpen
  \bibfield  {author} {\bibinfo {author} {\bibfnamefont {S.}~\bibnamefont {Chatrchyan}} \emph {et~al.} (\bibinfo {collaboration} {CMS}),\ }\href {\doibase 10.1016/j.physletb.2012.08.021} {\bibfield  {journal} {\bibinfo  {journal} {Phys. Lett. B}\ }\textbf {\bibinfo {volume} {716}},\ \bibinfo {pages} {30} (\bibinfo {year} {2012})},\ \Eprint {http://arxiv.org/abs/1207.7235} {arXiv:1207.7235 [hep-ex]} \BibitemShut {NoStop}%
\bibitem [{\citenamefont {Aad}\ \emph {et~al.}(2020)\citenamefont {Aad} \emph {et~al.}}]{ATLAS:2020qcv}%
  \BibitemOpen
  \bibfield  {author} {\bibinfo {author} {\bibfnamefont {G.}~\bibnamefont {Aad}} \emph {et~al.} (\bibinfo {collaboration} {ATLAS}),\ }\href {\doibase 10.1016/j.physletb.2020.135754} {\bibfield  {journal} {\bibinfo  {journal} {Phys. Lett. B}\ }\textbf {\bibinfo {volume} {809}},\ \bibinfo {pages} {135754} (\bibinfo {year} {2020})},\ \Eprint {http://arxiv.org/abs/2005.05382} {arXiv:2005.05382 [hep-ex]} \BibitemShut {NoStop}%
\bibitem [{\citenamefont {Tumasyan}\ \emph {et~al.}(2023)\citenamefont {Tumasyan} \emph {et~al.}}]{CMS:2022ahq}%
  \BibitemOpen
  \bibfield  {author} {\bibinfo {author} {\bibfnamefont {A.}~\bibnamefont {Tumasyan}} \emph {et~al.} (\bibinfo {collaboration} {CMS}),\ }\href {\doibase 10.1007/JHEP05(2023)233} {\bibfield  {journal} {\bibinfo  {journal} {JHEP}\ }\textbf {\bibinfo {volume} {05}},\ \bibinfo {pages} {233} (\bibinfo {year} {2023})},\ \Eprint {http://arxiv.org/abs/2204.12945} {arXiv:2204.12945 [hep-ex]} \BibitemShut {NoStop}%
\bibitem [{\citenamefont {Aad}\ \emph {et~al.}(2024)\citenamefont {Aad} \emph {et~al.}}]{ATLAS:2023yqk}%
  \BibitemOpen
  \bibfield  {author} {\bibinfo {author} {\bibfnamefont {G.}~\bibnamefont {Aad}} \emph {et~al.} (\bibinfo {collaboration} {ATLAS, CMS}),\ }\href {\doibase 10.1103/PhysRevLett.132.021803} {\bibfield  {journal} {\bibinfo  {journal} {Phys. Rev. Lett.}\ }\textbf {\bibinfo {volume} {132}},\ \bibinfo {pages} {021803} (\bibinfo {year} {2024})},\ \Eprint {http://arxiv.org/abs/2309.03501} {arXiv:2309.03501 [hep-ex]} \BibitemShut {NoStop}%
\bibitem [{\citenamefont {Barducci}\ \emph {et~al.}(2023)\citenamefont {Barducci}, \citenamefont {Di~Luzio}, \citenamefont {Nardecchia},\ and\ \citenamefont {Toni}}]{Barducci:2023zml}%
  \BibitemOpen
  \bibfield  {author} {\bibinfo {author} {\bibfnamefont {D.}~\bibnamefont {Barducci}}, \bibinfo {author} {\bibfnamefont {L.}~\bibnamefont {Di~Luzio}}, \bibinfo {author} {\bibfnamefont {M.}~\bibnamefont {Nardecchia}}, \ and\ \bibinfo {author} {\bibfnamefont {C.}~\bibnamefont {Toni}},\ }\href {\doibase 10.1007/JHEP12(2023)154} {\bibfield  {journal} {\bibinfo  {journal} {JHEP}\ }\textbf {\bibinfo {volume} {12}},\ \bibinfo {pages} {154} (\bibinfo {year} {2023})},\ \Eprint {http://arxiv.org/abs/2311.10130} {arXiv:2311.10130 [hep-ph]} \BibitemShut {NoStop}%
\bibitem [{\citenamefont {Boto}\ \emph {et~al.}(2024)\citenamefont {Boto}, \citenamefont {Das}, \citenamefont {Romao}, \citenamefont {Saha},\ and\ \citenamefont {Silva}}]{Boto:2023bpg}%
  \BibitemOpen
  \bibfield  {author} {\bibinfo {author} {\bibfnamefont {R.}~\bibnamefont {Boto}}, \bibinfo {author} {\bibfnamefont {D.}~\bibnamefont {Das}}, \bibinfo {author} {\bibfnamefont {J.~C.}\ \bibnamefont {Romao}}, \bibinfo {author} {\bibfnamefont {I.}~\bibnamefont {Saha}}, \ and\ \bibinfo {author} {\bibfnamefont {J.~P.}\ \bibnamefont {Silva}},\ }\href {\doibase 10.1103/PhysRevD.109.095002} {\bibfield  {journal} {\bibinfo  {journal} {Phys. Rev. D}\ }\textbf {\bibinfo {volume} {109}},\ \bibinfo {pages} {095002} (\bibinfo {year} {2024})},\ \Eprint {http://arxiv.org/abs/2312.13050} {arXiv:2312.13050 [hep-ph]} \BibitemShut {NoStop}%
\bibitem [{\citenamefont {Das}\ \emph {et~al.}(2024)\citenamefont {Das}, \citenamefont {Jha},\ and\ \citenamefont {Nanda}}]{Das:2024tfe}%
  \BibitemOpen
  \bibfield  {author} {\bibinfo {author} {\bibfnamefont {N.}~\bibnamefont {Das}}, \bibinfo {author} {\bibfnamefont {T.}~\bibnamefont {Jha}}, \ and\ \bibinfo {author} {\bibfnamefont {D.}~\bibnamefont {Nanda}},\ }\href {\doibase 10.1103/PhysRevD.109.115020} {\bibfield  {journal} {\bibinfo  {journal} {Phys. Rev. D}\ }\textbf {\bibinfo {volume} {109}},\ \bibinfo {pages} {115020} (\bibinfo {year} {2024})},\ \Eprint {http://arxiv.org/abs/2402.01317} {arXiv:2402.01317 [hep-ph]} \BibitemShut {NoStop}%
\bibitem [{\citenamefont {Cheung}\ and\ \citenamefont {Ouseph}(2024)}]{Cheung:2024kml}%
  \BibitemOpen
  \bibfield  {author} {\bibinfo {author} {\bibfnamefont {K.}~\bibnamefont {Cheung}}\ and\ \bibinfo {author} {\bibfnamefont {C.~J.}\ \bibnamefont {Ouseph}},\ }\href {\doibase 10.1103/PhysRevD.110.055016} {\bibfield  {journal} {\bibinfo  {journal} {Phys. Rev. D}\ }\textbf {\bibinfo {volume} {110}},\ \bibinfo {pages} {055016} (\bibinfo {year} {2024})},\ \Eprint {http://arxiv.org/abs/2402.05678} {arXiv:2402.05678 [hep-ph]} \BibitemShut {NoStop}%
\bibitem [{\citenamefont {Hu}\ \emph {et~al.}(2024)\citenamefont {Hu}, \citenamefont {Zhang}, \citenamefont {Zhu},\ and\ \citenamefont {Chen}}]{Hu:2024slu}%
  \BibitemOpen
  \bibfield  {author} {\bibinfo {author} {\bibfnamefont {H.-c.}\ \bibnamefont {Hu}}, \bibinfo {author} {\bibfnamefont {Z.-Y.}\ \bibnamefont {Zhang}}, \bibinfo {author} {\bibfnamefont {N.-Y.}\ \bibnamefont {Zhu}}, \ and\ \bibinfo {author} {\bibfnamefont {H.-X.}\ \bibnamefont {Chen}},\ }\href {\doibase 10.1088/1674-1137/ad5427} {\bibfield  {journal} {\bibinfo  {journal} {Chin. Phys. C}\ }\textbf {\bibinfo {volume} {48}},\ \bibinfo {pages} {093101} (\bibinfo {year} {2024})},\ \Eprint {http://arxiv.org/abs/2406.00946} {arXiv:2406.00946 [hep-ph]} \BibitemShut {NoStop}%
\bibitem [{\citenamefont {Hern\'andez-Ju\'arez}\ \emph {et~al.}(2024)\citenamefont {Hern\'andez-Ju\'arez}, \citenamefont {Gait\'an},\ and\ \citenamefont {Martinez}}]{Hernandez-Juarez:2024pty}%
  \BibitemOpen
  \bibfield  {author} {\bibinfo {author} {\bibfnamefont {A.~I.}\ \bibnamefont {Hern\'andez-Ju\'arez}}, \bibinfo {author} {\bibfnamefont {R.}~\bibnamefont {Gait\'an}}, \ and\ \bibinfo {author} {\bibfnamefont {R.}~\bibnamefont {Martinez}},\ }\href@noop {} {\  (\bibinfo {year} {2024})},\ \Eprint {http://arxiv.org/abs/2405.03094} {arXiv:2405.03094 [hep-ph]} \BibitemShut {NoStop}%
\bibitem [{\citenamefont {Zhang}\ \emph {et~al.}(2024)\citenamefont {Zhang}, \citenamefont {Yue}, \citenamefont {Sun},\ and\ \citenamefont {Wang}}]{Zhang:2024yqw}%
  \BibitemOpen
  \bibfield  {author} {\bibinfo {author} {\bibfnamefont {X.-M.}\ \bibnamefont {Zhang}}, \bibinfo {author} {\bibfnamefont {C.-X.}\ \bibnamefont {Yue}}, \bibinfo {author} {\bibfnamefont {X.-C.}\ \bibnamefont {Sun}}, \ and\ \bibinfo {author} {\bibfnamefont {Y.-Q.}\ \bibnamefont {Wang}},\ }\href {\doibase 10.1142/S0217751X24501124} {\bibfield  {journal} {\bibinfo  {journal} {Int. J. Mod. Phys. A}\ }\textbf {\bibinfo {volume} {39}},\ \bibinfo {pages} {2450112} (\bibinfo {year} {2024})}\BibitemShut {NoStop}%
\bibitem [{\citenamefont {Arhrib}\ \emph {et~al.}(2024)\citenamefont {Arhrib}, \citenamefont {Phan}, \citenamefont {Tran},\ and\ \citenamefont {Yuan}}]{Arhrib:2024wjj}%
  \BibitemOpen
  \bibfield  {author} {\bibinfo {author} {\bibfnamefont {A.}~\bibnamefont {Arhrib}}, \bibinfo {author} {\bibfnamefont {K.~H.}\ \bibnamefont {Phan}}, \bibinfo {author} {\bibfnamefont {V.~Q.}\ \bibnamefont {Tran}}, \ and\ \bibinfo {author} {\bibfnamefont {T.-C.}\ \bibnamefont {Yuan}},\ }\href@noop {} {\  (\bibinfo {year} {2024})},\ \Eprint {http://arxiv.org/abs/2405.03127} {arXiv:2405.03127 [hep-ph]} \BibitemShut {NoStop}%
\bibitem [{\citenamefont {Israr}\ and\ \citenamefont {Rehman}(2024)}]{Israr:2024ubp}%
  \BibitemOpen
  \bibfield  {author} {\bibinfo {author} {\bibfnamefont {S.}~\bibnamefont {Israr}}\ and\ \bibinfo {author} {\bibfnamefont {M.}~\bibnamefont {Rehman}},\ }\href@noop {} {\  (\bibinfo {year} {2024})},\ \Eprint {http://arxiv.org/abs/2407.01210} {arXiv:2407.01210 [hep-ph]} \BibitemShut {NoStop}%
\bibitem [{\citenamefont {Mantzaropoulos}(2024)}]{Mantzaropoulos:2024vpe}%
  \BibitemOpen
  \bibfield  {author} {\bibinfo {author} {\bibfnamefont {K.}~\bibnamefont {Mantzaropoulos}},\ }\href {\doibase 10.1103/PhysRevD.110.055041} {\bibfield  {journal} {\bibinfo  {journal} {Phys. Rev. D}\ }\textbf {\bibinfo {volume} {110}},\ \bibinfo {pages} {055041} (\bibinfo {year} {2024})},\ \Eprint {http://arxiv.org/abs/2407.09145} {arXiv:2407.09145 [hep-ph]} \BibitemShut {NoStop}%
\bibitem [{\citenamefont {Sang}\ \emph {et~al.}(2024)\citenamefont {Sang}, \citenamefont {Feng},\ and\ \citenamefont {Jia}}]{Sang:2024vqk}%
  \BibitemOpen
  \bibfield  {author} {\bibinfo {author} {\bibfnamefont {W.-L.}\ \bibnamefont {Sang}}, \bibinfo {author} {\bibfnamefont {F.}~\bibnamefont {Feng}}, \ and\ \bibinfo {author} {\bibfnamefont {Y.}~\bibnamefont {Jia}},\ }\href {\doibase 10.1103/PhysRevD.110.L051302} {\bibfield  {journal} {\bibinfo  {journal} {Phys. Rev. D}\ }\textbf {\bibinfo {volume} {110}},\ \bibinfo {pages} {L051302} (\bibinfo {year} {2024})},\ \Eprint {http://arxiv.org/abs/2405.03464} {arXiv:2405.03464 [hep-ph]} \BibitemShut {NoStop}%
\bibitem [{\citenamefont {Kachanovich}\ \emph {et~al.}(2020)\citenamefont {Kachanovich}, \citenamefont {Nierste},\ and\ \citenamefont {Ni\v{s}and\v{z}i\'c}}]{Kachanovich:2020xyg}%
  \BibitemOpen
  \bibfield  {author} {\bibinfo {author} {\bibfnamefont {A.}~\bibnamefont {Kachanovich}}, \bibinfo {author} {\bibfnamefont {U.}~\bibnamefont {Nierste}}, \ and\ \bibinfo {author} {\bibfnamefont {I.}~\bibnamefont {Ni\v{s}and\v{z}i\'c}},\ }\href {\doibase 10.1103/PhysRevD.101.073003} {\bibfield  {journal} {\bibinfo  {journal} {Phys. Rev. D}\ }\textbf {\bibinfo {volume} {101}},\ \bibinfo {pages} {073003} (\bibinfo {year} {2020})},\ \Eprint {http://arxiv.org/abs/2001.06516} {arXiv:2001.06516 [hep-ph]} \BibitemShut {NoStop}%
\bibitem [{\citenamefont {Kachanovich}\ \emph {et~al.}(2022)\citenamefont {Kachanovich}, \citenamefont {Nierste},\ and\ \citenamefont {Ni\v{s}and\v{z}i\'c}}]{Kachanovich:2021pvx}%
  \BibitemOpen
  \bibfield  {author} {\bibinfo {author} {\bibfnamefont {A.}~\bibnamefont {Kachanovich}}, \bibinfo {author} {\bibfnamefont {U.}~\bibnamefont {Nierste}}, \ and\ \bibinfo {author} {\bibfnamefont {I.}~\bibnamefont {Ni\v{s}and\v{z}i\'c}},\ }\href {\doibase 10.1103/PhysRevD.105.013007} {\bibfield  {journal} {\bibinfo  {journal} {Phys. Rev. D}\ }\textbf {\bibinfo {volume} {105}},\ \bibinfo {pages} {013007} (\bibinfo {year} {2022})},\ \Eprint {http://arxiv.org/abs/2109.04426} {arXiv:2109.04426 [hep-ph]} \BibitemShut {NoStop}%
\bibitem [{\citenamefont {Corbett}\ and\ \citenamefont {Rasmussen}(2022)}]{Corbett:2021iob}%
  \BibitemOpen
  \bibfield  {author} {\bibinfo {author} {\bibfnamefont {T.}~\bibnamefont {Corbett}}\ and\ \bibinfo {author} {\bibfnamefont {T.}~\bibnamefont {Rasmussen}},\ }\href {\doibase 10.21468/SciPostPhys.13.5.112} {\bibfield  {journal} {\bibinfo  {journal} {SciPost Phys.}\ }\textbf {\bibinfo {volume} {13}},\ \bibinfo {pages} {112} (\bibinfo {year} {2022})},\ \Eprint {http://arxiv.org/abs/2110.03694} {arXiv:2110.03694 [hep-ph]} \BibitemShut {NoStop}%
\bibitem [{\citenamefont {Chen}\ \emph {et~al.}(2022)\citenamefont {Chen}, \citenamefont {Gehrmann}, \citenamefont {Glover},\ and\ \citenamefont {Huss}}]{Chen:2021ibm}%
  \BibitemOpen
  \bibfield  {author} {\bibinfo {author} {\bibfnamefont {X.}~\bibnamefont {Chen}}, \bibinfo {author} {\bibfnamefont {T.}~\bibnamefont {Gehrmann}}, \bibinfo {author} {\bibfnamefont {E.~W.~N.}\ \bibnamefont {Glover}}, \ and\ \bibinfo {author} {\bibfnamefont {A.}~\bibnamefont {Huss}},\ }\href {\doibase 10.1007/JHEP01(2022)053} {\bibfield  {journal} {\bibinfo  {journal} {JHEP}\ }\textbf {\bibinfo {volume} {01}},\ \bibinfo {pages} {053} (\bibinfo {year} {2022})},\ \Eprint {http://arxiv.org/abs/2111.02157} {arXiv:2111.02157 [hep-ph]} \BibitemShut {NoStop}%
\bibitem [{\citenamefont {Ahmed}\ \emph {et~al.}(2024)\citenamefont {Ahmed}, \citenamefont {Hasan}, \citenamefont {Iqbal}, \citenamefont {Junaid}, \citenamefont {Tariq},\ and\ \citenamefont {Uzair}}]{Ahmed:2023vyl}%
  \BibitemOpen
  \bibfield  {author} {\bibinfo {author} {\bibfnamefont {I.}~\bibnamefont {Ahmed}}, \bibinfo {author} {\bibfnamefont {U.}~\bibnamefont {Hasan}}, \bibinfo {author} {\bibfnamefont {S.}~\bibnamefont {Iqbal}}, \bibinfo {author} {\bibfnamefont {M.}~\bibnamefont {Junaid}}, \bibinfo {author} {\bibfnamefont {B.}~\bibnamefont {Tariq}}, \ and\ \bibinfo {author} {\bibfnamefont {A.}~\bibnamefont {Uzair}},\ }\href {\doibase 10.1007/JHEP05(2024)187} {\bibfield  {journal} {\bibinfo  {journal} {JHEP}\ }\textbf {\bibinfo {volume} {05}},\ \bibinfo {pages} {187} (\bibinfo {year} {2024})},\ \Eprint {http://arxiv.org/abs/2309.07448} {arXiv:2309.07448 [hep-ph]} \BibitemShut {NoStop}%
\bibitem [{\citenamefont {Hue}\ \emph {et~al.}(2023)\citenamefont {Hue}, \citenamefont {Tran}, \citenamefont {Nguyen},\ and\ \citenamefont {Phan}}]{Hue:2023tdz}%
  \BibitemOpen
  \bibfield  {author} {\bibinfo {author} {\bibfnamefont {L.~T.}\ \bibnamefont {Hue}}, \bibinfo {author} {\bibfnamefont {D.~T.}\ \bibnamefont {Tran}}, \bibinfo {author} {\bibfnamefont {T.~H.}\ \bibnamefont {Nguyen}}, \ and\ \bibinfo {author} {\bibfnamefont {K.~H.}\ \bibnamefont {Phan}},\ }\href {\doibase 10.1093/ptep/ptad106} {\bibfield  {journal} {\bibinfo  {journal} {PTEP}\ }\textbf {\bibinfo {volume} {2023}},\ \bibinfo {pages} {083B06} (\bibinfo {year} {2023})},\ \Eprint {http://arxiv.org/abs/2305.04002} {arXiv:2305.04002 [hep-ph]} \BibitemShut {NoStop}%
\bibitem [{\citenamefont {Van~On}\ \emph {et~al.}(2022)\citenamefont {Van~On}, \citenamefont {Tran}, \citenamefont {Nguyen},\ and\ \citenamefont {Phan}}]{VanOn:2021myp}%
  \BibitemOpen
  \bibfield  {author} {\bibinfo {author} {\bibfnamefont {V.}~\bibnamefont {Van~On}}, \bibinfo {author} {\bibfnamefont {D.~T.}\ \bibnamefont {Tran}}, \bibinfo {author} {\bibfnamefont {C.~L.}\ \bibnamefont {Nguyen}}, \ and\ \bibinfo {author} {\bibfnamefont {K.~H.}\ \bibnamefont {Phan}},\ }\href {\doibase 10.1140/epjc/s10052-022-10225-z} {\bibfield  {journal} {\bibinfo  {journal} {Eur. Phys. J. C}\ }\textbf {\bibinfo {volume} {82}},\ \bibinfo {pages} {277} (\bibinfo {year} {2022})},\ \Eprint {http://arxiv.org/abs/2111.07708} {arXiv:2111.07708 [hep-ph]} \BibitemShut {NoStop}%
\bibitem [{\citenamefont {Sun}\ and\ \citenamefont {Gao}(2014)}]{Sun:2013cba}%
  \BibitemOpen
  \bibfield  {author} {\bibinfo {author} {\bibfnamefont {Y.}~\bibnamefont {Sun}}\ and\ \bibinfo {author} {\bibfnamefont {D.-N.}\ \bibnamefont {Gao}},\ }\href {\doibase 10.1103/PhysRevD.89.017301} {\bibfield  {journal} {\bibinfo  {journal} {Phys. Rev. D}\ }\textbf {\bibinfo {volume} {89}},\ \bibinfo {pages} {017301} (\bibinfo {year} {2014})},\ \Eprint {http://arxiv.org/abs/1310.8404} {arXiv:1310.8404 [hep-ph]} \BibitemShut {NoStop}%
\bibitem [{\citenamefont {Phan}\ \emph {et~al.}(2021)\citenamefont {Phan}, \citenamefont {Hue},\ and\ \citenamefont {Tran}}]{Phan:2021xwc}%
  \BibitemOpen
  \bibfield  {author} {\bibinfo {author} {\bibfnamefont {K.~H.}\ \bibnamefont {Phan}}, \bibinfo {author} {\bibfnamefont {L.~T.}\ \bibnamefont {Hue}}, \ and\ \bibinfo {author} {\bibfnamefont {D.~T.}\ \bibnamefont {Tran}},\ }\href {\doibase 10.1093/ptep/ptab121} {\bibfield  {journal} {\bibinfo  {journal} {PTEP}\ }\textbf {\bibinfo {volume} {2021}},\ \bibinfo {pages} {103B07} (\bibinfo {year} {2021})},\ \Eprint {http://arxiv.org/abs/2106.14466} {arXiv:2106.14466 [hep-ph]} \BibitemShut {NoStop}%
\bibitem [{\citenamefont {Phan}\ and\ \citenamefont {Tran}(2022)}]{Phan:2021ovj}%
  \BibitemOpen
  \bibfield  {author} {\bibinfo {author} {\bibfnamefont {K.~H.}\ \bibnamefont {Phan}}\ and\ \bibinfo {author} {\bibfnamefont {D.~T.}\ \bibnamefont {Tran}},\ }\href {\doibase 10.1093/ptep/ptac012} {\bibfield  {journal} {\bibinfo  {journal} {PTEP}\ }\textbf {\bibinfo {volume} {2022}},\ \bibinfo {pages} {023B03} (\bibinfo {year} {2022})},\ \Eprint {http://arxiv.org/abs/2111.07698} {arXiv:2111.07698 [hep-ph]} \BibitemShut {NoStop}%
\bibitem [{\citenamefont {Kachanovich}\ and\ \citenamefont {Ni\v{s}and\v{z}i\'c}(2024)}]{Kachanovich:2024vpt}%
  \BibitemOpen
  \bibfield  {author} {\bibinfo {author} {\bibfnamefont {A.}~\bibnamefont {Kachanovich}}\ and\ \bibinfo {author} {\bibfnamefont {I.}~\bibnamefont {Ni\v{s}and\v{z}i\'c}},\ }\href@noop {} {\  (\bibinfo {year} {2024})},\ \Eprint {http://arxiv.org/abs/2405.16239} {arXiv:2405.16239 [hep-ph]} \BibitemShut {NoStop}%
\bibitem [{\citenamefont {Abbasabadi}\ \emph {et~al.}(1997)\citenamefont {Abbasabadi}, \citenamefont {Bowser-Chao}, \citenamefont {Dicus},\ and\ \citenamefont {Repko}}]{Abbasabadi:1996ze}%
  \BibitemOpen
  \bibfield  {author} {\bibinfo {author} {\bibfnamefont {A.}~\bibnamefont {Abbasabadi}}, \bibinfo {author} {\bibfnamefont {D.}~\bibnamefont {Bowser-Chao}}, \bibinfo {author} {\bibfnamefont {D.~A.}\ \bibnamefont {Dicus}}, \ and\ \bibinfo {author} {\bibfnamefont {W.~W.}\ \bibnamefont {Repko}},\ }\href {\doibase 10.1103/PhysRevD.55.5647} {\bibfield  {journal} {\bibinfo  {journal} {Phys. Rev. D}\ }\textbf {\bibinfo {volume} {55}},\ \bibinfo {pages} {5647} (\bibinfo {year} {1997})},\ \Eprint {http://arxiv.org/abs/hep-ph/9611209} {arXiv:hep-ph/9611209} \BibitemShut {NoStop}%
\bibitem [{\citenamefont {Chen}\ \emph {et~al.}(2013)\citenamefont {Chen}, \citenamefont {Qiao},\ and\ \citenamefont {Zhu}}]{Chen:2012ju}%
  \BibitemOpen
  \bibfield  {author} {\bibinfo {author} {\bibfnamefont {L.-B.}\ \bibnamefont {Chen}}, \bibinfo {author} {\bibfnamefont {C.-F.}\ \bibnamefont {Qiao}}, \ and\ \bibinfo {author} {\bibfnamefont {R.-L.}\ \bibnamefont {Zhu}},\ }\href {\doibase 10.1016/j.physletb.2013.08.050} {\bibfield  {journal} {\bibinfo  {journal} {Phys. Lett. B}\ }\textbf {\bibinfo {volume} {726}},\ \bibinfo {pages} {306} (\bibinfo {year} {2013})},\ \bibinfo {note} {[Erratum: Phys.Lett.B 808, 135629 (2020)]},\ \Eprint {http://arxiv.org/abs/1211.6058} {arXiv:1211.6058 [hep-ph]} \BibitemShut {NoStop}%
\bibitem [{\citenamefont {Dicus}\ and\ \citenamefont {Repko}(2013)}]{Dicus:2013ycd}%
  \BibitemOpen
  \bibfield  {author} {\bibinfo {author} {\bibfnamefont {D.~A.}\ \bibnamefont {Dicus}}\ and\ \bibinfo {author} {\bibfnamefont {W.~W.}\ \bibnamefont {Repko}},\ }\href {\doibase 10.1103/PhysRevD.87.077301} {\bibfield  {journal} {\bibinfo  {journal} {Phys. Rev. D}\ }\textbf {\bibinfo {volume} {87}},\ \bibinfo {pages} {077301} (\bibinfo {year} {2013})},\ \Eprint {http://arxiv.org/abs/1302.2159} {arXiv:1302.2159 [hep-ph]} \BibitemShut {NoStop}%
\bibitem [{\citenamefont {Passarino}(2013)}]{Passarino:2013nka}%
  \BibitemOpen
  \bibfield  {author} {\bibinfo {author} {\bibfnamefont {G.}~\bibnamefont {Passarino}},\ }\href {\doibase 10.1016/j.physletb.2013.10.052} {\bibfield  {journal} {\bibinfo  {journal} {Phys. Lett. B}\ }\textbf {\bibinfo {volume} {727}},\ \bibinfo {pages} {424} (\bibinfo {year} {2013})},\ \Eprint {http://arxiv.org/abs/1308.0422} {arXiv:1308.0422 [hep-ph]} \BibitemShut {NoStop}%
\bibitem [{\citenamefont {Han}\ and\ \citenamefont {Wang}(2017)}]{Han:2017yhy}%
  \BibitemOpen
  \bibfield  {author} {\bibinfo {author} {\bibfnamefont {T.}~\bibnamefont {Han}}\ and\ \bibinfo {author} {\bibfnamefont {X.}~\bibnamefont {Wang}},\ }\href {\doibase 10.1007/JHEP10(2017)036} {\bibfield  {journal} {\bibinfo  {journal} {JHEP}\ }\textbf {\bibinfo {volume} {10}},\ \bibinfo {pages} {036} (\bibinfo {year} {2017})},\ \Eprint {http://arxiv.org/abs/1704.00790} {arXiv:1704.00790 [hep-ph]} \BibitemShut {NoStop}%
\bibitem [{\citenamefont {Ni{\v{s}}and{\v{z}}i{\'c}}(2025)}]{Nisandzic:2025raw}%
  \BibitemOpen
  \bibfield  {author} {\bibinfo {author} {\bibfnamefont {I.}~\bibnamefont {Ni{\v{s}}and{\v{z}}i{\'c}}},\ }\href {\doibase 10.22323/1.481.0009} {\bibfield  {journal} {\bibinfo  {journal} {PoS}\ }\textbf {\bibinfo {volume} {DISCRETE2024}},\ \bibinfo {pages} {009} (\bibinfo {year} {2025})}\BibitemShut {NoStop}%
\bibitem [{\citenamefont {Kachanovich}\ \emph {et~al.}(2025)\citenamefont {Kachanovich}, \citenamefont {Kimus}, \citenamefont {Lowette},\ and\ \citenamefont {Tytgat}}]{Kachanovich:2025cxz}%
  \BibitemOpen
  \bibfield  {author} {\bibinfo {author} {\bibfnamefont {A.}~\bibnamefont {Kachanovich}}, \bibinfo {author} {\bibfnamefont {J.}~\bibnamefont {Kimus}}, \bibinfo {author} {\bibfnamefont {S.}~\bibnamefont {Lowette}}, \ and\ \bibinfo {author} {\bibfnamefont {M.~H.~G.}\ \bibnamefont {Tytgat}},\ }\href@noop {} {\  (\bibinfo {year} {2025})},\ \Eprint {http://arxiv.org/abs/2503.08659} {arXiv:2503.08659 [hep-ph]} \BibitemShut {NoStop}%
\bibitem [{\citenamefont {Aad}\ \emph {et~al.}(2025)\citenamefont {Aad} \emph {et~al.}}]{ATLAS:2025aip}%
  \BibitemOpen
  \bibfield  {author} {\bibinfo {author} {\bibfnamefont {G.}~\bibnamefont {Aad}} \emph {et~al.} (\bibinfo {collaboration} {ATLAS}),\ }\href@noop {} {\  (\bibinfo {year} {2025})},\ \Eprint {http://arxiv.org/abs/2507.12598} {arXiv:2507.12598 [hep-ex]} \BibitemShut {NoStop}%
\bibitem [{\citenamefont {Kachanovich}(2021)}]{Kachanovich:2020dah}%
  \BibitemOpen
  \bibfield  {author} {\bibinfo {author} {\bibfnamefont {A.}~\bibnamefont {Kachanovich}},\ }\emph {\bibinfo {title} {{Flavour-changing neutral current processes beyond the Standard Model}}},\ \href {\doibase 10.5445/IR/1000130547} {Ph.D. thesis},\ \bibinfo  {school} {KIT, Karlsruhe, Dept. Phys.} (\bibinfo {year} {2021})\BibitemShut {NoStop}%
\bibitem [{\citenamefont {Toma}(2013)}]{Toma:2013bka}%
  \BibitemOpen
  \bibfield  {author} {\bibinfo {author} {\bibfnamefont {T.}~\bibnamefont {Toma}},\ }\href {\doibase 10.1103/PhysRevLett.111.091301} {\bibfield  {journal} {\bibinfo  {journal} {Phys. Rev. Lett.}\ }\textbf {\bibinfo {volume} {111}},\ \bibinfo {pages} {091301} (\bibinfo {year} {2013})},\ \Eprint {http://arxiv.org/abs/1307.6181} {arXiv:1307.6181 [hep-ph]} \BibitemShut {NoStop}%
\bibitem [{\citenamefont {Giacchino}\ \emph {et~al.}(2013)\citenamefont {Giacchino}, \citenamefont {Lopez-Honorez},\ and\ \citenamefont {Tytgat}}]{Giacchino:2013bta}%
  \BibitemOpen
  \bibfield  {author} {\bibinfo {author} {\bibfnamefont {F.}~\bibnamefont {Giacchino}}, \bibinfo {author} {\bibfnamefont {L.}~\bibnamefont {Lopez-Honorez}}, \ and\ \bibinfo {author} {\bibfnamefont {M.~H.~G.}\ \bibnamefont {Tytgat}},\ }\href {\doibase 10.1088/1475-7516/2013/10/025} {\bibfield  {journal} {\bibinfo  {journal} {JCAP}\ }\textbf {\bibinfo {volume} {10}},\ \bibinfo {pages} {025} (\bibinfo {year} {2013})},\ \Eprint {http://arxiv.org/abs/1307.6480} {arXiv:1307.6480 [hep-ph]} \BibitemShut {NoStop}%
\bibitem [{\citenamefont {Silveira}\ and\ \citenamefont {Zee}(1985)}]{Silveira:1985rk}%
  \BibitemOpen
  \bibfield  {author} {\bibinfo {author} {\bibfnamefont {V.}~\bibnamefont {Silveira}}\ and\ \bibinfo {author} {\bibfnamefont {A.}~\bibnamefont {Zee}},\ }\href {\doibase 10.1016/0370-2693(85)90624-0} {\bibfield  {journal} {\bibinfo  {journal} {Phys. Lett. B}\ }\textbf {\bibinfo {volume} {161}},\ \bibinfo {pages} {136} (\bibinfo {year} {1985})}\BibitemShut {NoStop}%
\bibitem [{\citenamefont {McDonald}(1994)}]{McDonald:1993ex}%
  \BibitemOpen
  \bibfield  {author} {\bibinfo {author} {\bibfnamefont {J.}~\bibnamefont {McDonald}},\ }\href {\doibase 10.1103/PhysRevD.50.3637} {\bibfield  {journal} {\bibinfo  {journal} {Phys. Rev. D}\ }\textbf {\bibinfo {volume} {50}},\ \bibinfo {pages} {3637} (\bibinfo {year} {1994})},\ \Eprint {http://arxiv.org/abs/hep-ph/0702143} {arXiv:hep-ph/0702143} \BibitemShut {NoStop}%
\bibitem [{\citenamefont {Burgess}\ \emph {et~al.}(2001)\citenamefont {Burgess}, \citenamefont {Pospelov},\ and\ \citenamefont {ter Veldhuis}}]{Burgess:2000yq}%
  \BibitemOpen
  \bibfield  {author} {\bibinfo {author} {\bibfnamefont {C.~P.}\ \bibnamefont {Burgess}}, \bibinfo {author} {\bibfnamefont {M.}~\bibnamefont {Pospelov}}, \ and\ \bibinfo {author} {\bibfnamefont {T.}~\bibnamefont {ter Veldhuis}},\ }\href {\doibase 10.1016/S0550-3213(01)00513-2} {\bibfield  {journal} {\bibinfo  {journal} {Nucl. Phys. B}\ }\textbf {\bibinfo {volume} {619}},\ \bibinfo {pages} {709} (\bibinfo {year} {2001})},\ \Eprint {http://arxiv.org/abs/hep-ph/0011335} {arXiv:hep-ph/0011335} \BibitemShut {NoStop}%
\bibitem [{\citenamefont {Shtabovenko}\ \emph {et~al.}(2016)\citenamefont {Shtabovenko}, \citenamefont {Mertig},\ and\ \citenamefont {Orellana}}]{Shtabovenko:2016sxi}%
  \BibitemOpen
  \bibfield  {author} {\bibinfo {author} {\bibfnamefont {V.}~\bibnamefont {Shtabovenko}}, \bibinfo {author} {\bibfnamefont {R.}~\bibnamefont {Mertig}}, \ and\ \bibinfo {author} {\bibfnamefont {F.}~\bibnamefont {Orellana}},\ }\href {\doibase 10.1016/j.cpc.2016.06.008} {\bibfield  {journal} {\bibinfo  {journal} {Comput. Phys. Commun.}\ }\textbf {\bibinfo {volume} {207}},\ \bibinfo {pages} {432} (\bibinfo {year} {2016})},\ \Eprint {http://arxiv.org/abs/1601.01167} {arXiv:1601.01167 [hep-ph]} \BibitemShut {NoStop}%
\bibitem [{\citenamefont {Shtabovenko}\ \emph {et~al.}(2020)\citenamefont {Shtabovenko}, \citenamefont {Mertig},\ and\ \citenamefont {Orellana}}]{Shtabovenko:2020gxv}%
  \BibitemOpen
  \bibfield  {author} {\bibinfo {author} {\bibfnamefont {V.}~\bibnamefont {Shtabovenko}}, \bibinfo {author} {\bibfnamefont {R.}~\bibnamefont {Mertig}}, \ and\ \bibinfo {author} {\bibfnamefont {F.}~\bibnamefont {Orellana}},\ }\href {\doibase 10.1016/j.cpc.2020.107478} {\bibfield  {journal} {\bibinfo  {journal} {Comput. Phys. Commun.}\ }\textbf {\bibinfo {volume} {256}},\ \bibinfo {pages} {107478} (\bibinfo {year} {2020})},\ \Eprint {http://arxiv.org/abs/2001.04407} {arXiv:2001.04407 [hep-ph]} \BibitemShut {NoStop}%
\bibitem [{\citenamefont {Shtabovenko}\ \emph {et~al.}(2023)\citenamefont {Shtabovenko}, \citenamefont {Mertig},\ and\ \citenamefont {Orellana}}]{Shtabovenko:2023idz}%
  \BibitemOpen
  \bibfield  {author} {\bibinfo {author} {\bibfnamefont {V.}~\bibnamefont {Shtabovenko}}, \bibinfo {author} {\bibfnamefont {R.}~\bibnamefont {Mertig}}, \ and\ \bibinfo {author} {\bibfnamefont {F.}~\bibnamefont {Orellana}},\ }\href@noop {} {\  (\bibinfo {year} {2023})},\ \Eprint {http://arxiv.org/abs/2312.14089} {arXiv:2312.14089 [hep-ph]} \BibitemShut {NoStop}%
\bibitem [{\citenamefont {Aliberti}\ \emph {et~al.}(2025)\citenamefont {Aliberti} \emph {et~al.}}]{Aliberti:2025beg}%
  \BibitemOpen
  \bibfield  {author} {\bibinfo {author} {\bibfnamefont {R.}~\bibnamefont {Aliberti}} \emph {et~al.},\ }\href {\doibase 10.1016/j.physrep.2025.08.002} {\bibfield  {journal} {\bibinfo  {journal} {Phys. Rept.}\ }\textbf {\bibinfo {volume} {1143}},\ \bibinfo {pages} {1} (\bibinfo {year} {2025})},\ \Eprint {http://arxiv.org/abs/2505.21476} {arXiv:2505.21476 [hep-ph]} \BibitemShut {NoStop}%
\bibitem [{\citenamefont {Peskin}\ and\ \citenamefont {Takeuchi}(1990)}]{Peskin:1990zt}%
  \BibitemOpen
  \bibfield  {author} {\bibinfo {author} {\bibfnamefont {M.~E.}\ \bibnamefont {Peskin}}\ and\ \bibinfo {author} {\bibfnamefont {T.}~\bibnamefont {Takeuchi}},\ }\href {\doibase 10.1103/PhysRevLett.65.964} {\bibfield  {journal} {\bibinfo  {journal} {Phys. Rev. Lett.}\ }\textbf {\bibinfo {volume} {65}},\ \bibinfo {pages} {964} (\bibinfo {year} {1990})}\BibitemShut {NoStop}%
\bibitem [{\citenamefont {Maksymyk}\ \emph {et~al.}(1994)\citenamefont {Maksymyk}, \citenamefont {Burgess},\ and\ \citenamefont {London}}]{Maksymyk:1993zm}%
  \BibitemOpen
  \bibfield  {author} {\bibinfo {author} {\bibfnamefont {I.}~\bibnamefont {Maksymyk}}, \bibinfo {author} {\bibfnamefont {C.~P.}\ \bibnamefont {Burgess}}, \ and\ \bibinfo {author} {\bibfnamefont {D.}~\bibnamefont {London}},\ }\href {\doibase 10.1103/PhysRevD.50.529} {\bibfield  {journal} {\bibinfo  {journal} {Phys. Rev. D}\ }\textbf {\bibinfo {volume} {50}},\ \bibinfo {pages} {529} (\bibinfo {year} {1994})},\ \Eprint {http://arxiv.org/abs/hep-ph/9306267} {arXiv:hep-ph/9306267} \BibitemShut {NoStop}%
\bibitem [{\citenamefont {Albergaria}\ \emph {et~al.}(2024)\citenamefont {Albergaria}, \citenamefont {Jur\v{c}iukonis},\ and\ \citenamefont {Lavoura}}]{Albergaria:2023nby}%
  \BibitemOpen
  \bibfield  {author} {\bibinfo {author} {\bibfnamefont {F.}~\bibnamefont {Albergaria}}, \bibinfo {author} {\bibfnamefont {D.}~\bibnamefont {Jur\v{c}iukonis}}, \ and\ \bibinfo {author} {\bibfnamefont {L.}~\bibnamefont {Lavoura}},\ }\href {\doibase 10.1007/JHEP05(2024)190} {\bibfield  {journal} {\bibinfo  {journal} {JHEP}\ }\textbf {\bibinfo {volume} {05}},\ \bibinfo {pages} {190} (\bibinfo {year} {2024})},\ \Eprint {http://arxiv.org/abs/2312.09099} {arXiv:2312.09099 [hep-ph]} \BibitemShut {NoStop}%
\bibitem [{\citenamefont {Moreno}\ and\ \citenamefont {Tytgat}(1992)}]{Moreno:1991mf}%
  \BibitemOpen
  \bibfield  {author} {\bibinfo {author} {\bibfnamefont {J.~M.}\ \bibnamefont {Moreno}}\ and\ \bibinfo {author} {\bibfnamefont {M.}~\bibnamefont {Tytgat}},\ }\href {\doibase 10.1007/BF01558303} {\bibfield  {journal} {\bibinfo  {journal} {Z. Phys. C}\ }\textbf {\bibinfo {volume} {55}},\ \bibinfo {pages} {175} (\bibinfo {year} {1992})}\BibitemShut {NoStop}%
\end{thebibliography}%



\end{document}